# Stimulus dependence of the collective vibration of atoms in an icosahedral cluster

H. H. Liu, E. Y. Jiang[*], H. L. Bai, P. Wu, Z. Q. Li

Tianjin Key Laboratory of Low Dimensional Materials Physics and Preparing Technology and the Institute of Advanced Materials Physics, Faculty of Science, Tianjin University, No. 92, Weijin Road, Nankai district, Tianjin 300072, China

Chang Q Sun

School of Electrical and electronic Engineering, Nanyang Technological University, Singapore 639798, Singapore

Molecular dynamics calculations of the vibrational behavior of atoms in a Lennard-Jones 147-atom cluster revealed that the relaxation and the stability of the collective vibration of atoms in the cluster depend on the extent of the mechanical disturbance. A relatively larger-scale perturbation will cause a faster decay of the vibration magnitude, the potential and the kinetic energy compared to the vibration stimulated by a small-scale stimulus.

- Electronic mail: eyjiang@tju.edu.cn; ecqsun@ntu.edu.sg

# I. INTRODUCTION

Excitation by means of structure relaxation and electronic oscillation is of great importance to the understanding of the process of phase transition and energy transmission in materials.[1,2,3,4,5] Among all the possible modes of vibration, the collective vibration, or say collective excitation, stands out itself by consistent movement of a collection of particles.[6,7,8,9] Such collectively excited states were first discovered in nuclei.[10,11] The collective vibration was predicted to exist in the free electron gas in metals by Pines *et al.,*[12,13] and was then confirmed experimentally by Kaplan *et al.*[14] Using molecular dynamics simulation, Salian *et al,*[15] investigated the collective excitations in an $Ar_{13}$ cluster at different temperatures. They moved the twelve atoms on the surface radially away of the center atom by the same distance, causing the monopole cluster expanding to its maximum. They found in the subsequent relaxing process that the cluster gave radial uniform expansion and contraction, which is considered to be collective atomic vibration. The decay speed of the collective vibration depends on the temperature of operation.

In the present work, we introduced different dislocation stimulus into the icodahedral cluster containing 147-atom ($LJ_{147}$ cluster) with Lennard-Jones interatomic potential to study the relaxing process of different layers without repeating the temperature dependence. Different from the work conducted by Salian *et al.,*[15] we disturbed a certain portion of the atoms randomly in one direction. The monopole collective atomic vibration is found to be followed by a decay of the vibration. Results show that a stronger stimulus can accelerate the decaying of the collective vibration.

# II. METHOD

We may divide the LJ$_{147}$ cluster, from the central atom labeled as core-0 to the surface, into shell-1, shell-2, shell-3[16] as illustrated in Fig. 1. In order to show the relationship between the atoms clearly, we mark the atoms by groups [Fig. 1b]: the atoms on the vertex are marked as [*shell no*]a, those at the edges are marked as [*shell no*]b, those on the facets are marked as [*shell no*]c. For example, atoms on the vertex of shell-3 are marked as 3a; atoms at the edges of shell-2 are marked as 2b. Because there is only one sort of atoms in shell-1, we mark them as 1a.

The atoms in the LJ$_{147}$ cluster are bonded through the Lennard-Jones potential

$$P(r_{ij}) = 4\varepsilon\left[(\sigma/r_{ij})^{12} - (\sigma/r_{ij})^{6}\right], \quad (1)$$

where $r_{ij}$ is the distance between atom $i$ and $j$, $\varepsilon$ and $\sigma$ are used as the unit of energy and the characteristic length, respectively. Then the unit of temperature is $\varepsilon/k_B$ and the time is $\sqrt{m\sigma^2/\varepsilon}$ with $k_B$ being the Boltzmann constant and $m$ the atomic mass. We can obtain the positions and velocities of atoms each time using the Verlet algorithm,[17] the time step is set to be $10^{-4}\sqrt{m\sigma^2/\varepsilon}$.

At the beginning of the simulation, we introduced a weak and local stimulus into the LJ$_{147}$ cluster by moving randomly the specified atoms a short distance $d_{xyz}$ along the x, y and z direction. A larger $d_{xyz}$ value means a heavier stimulus. For convenience, we call $d_{xyz}$ as the scale of stimulus. Since the icosahedral LJ$_{147}$ cluster is the global optimized configuration for clusters of this size, introducing such a stimulus will provide excessive energy to the cluster. Hence, we can examine how the positions and energies of the atoms change during the excessive energy is released from the structure. We move randomly the atoms in core-2 (atoms in core-0, shell-1 and shell-2) and core-3 (atoms in core-0, shell-1, shell-2 and shell-3, or say all atoms of the entire cluster) by $d_{xyz}$, then the positions and energies of atoms in the LJ$_{147}$ cluster will change and decay till their original values.

## III. RESULTS

First, we introduced d$_{xyz}$ amounted at 0.01, 0.02, 0.03 and 0.04σ, into core-2. The distance D$^i$ between the ith atom and the cluster center, and the potential energy and the kinetic energy $E_k^i$ of each atom, are investigated during the excessive energy is released. The first $2\times10^6$ time steps of the relaxing process are recorded. To investigate the relaxing process in detail, we focused on the initial stage ($1\sim1\times10^5$ time step) and the final stage ($1.9\times10^6 \sim 2.0\times10^6$ time step).

Fig. 2 shows how the $E_k^i$ changes in the initial and final stages of the relaxation process, in which the scale of the stimulus $d_{xyz}$ is denoted. Fig. 2 shows that when the $d_{xyz}$ is small ($d_{xyz} = 0.01$), the fluctuation of the $E_k^i$ is weak. The fluctuation becomes stronger as the $d_{xyz}$ is increased (for instances, $d_{xyz} = 0.02$, 0.03, and 0.04), indicating that the atoms will gain more energy from stronger stimulus during the structure relaxation. From Fig. 2, we can also see that at the initial stage of relaxation, $E_k^i$ shows synchronous fluctuation. However, the synchronous status decays gradually as the relaxing process goes on, and which becomes faster when the $d_{xyz}$ increases. It can be seen from Fig. 2 that at the final stage of relaxation, when $d_{xyz} = 0.01$, the fluctuation of the $E_k^i$ still exhibits weak synchronous behavior, which could not be noticed when the $d_{xyz}$ is larger.

Fig. 3 shows the fluctuation of $D^i$ during the relaxation. It could be seen that the fluctuating curves of $D^i$ involve into groups according to the division in Fig. 1, which are denoted in Fig. 3. At the initial stage of the relaxation, the coincident fluctuation indicates clearly the collective movement of all the atoms. It can be found in Fig. 3 that the synchronous fluctuation of $D^i$ decays to chaotic status gradually as the relaxing process is going on, which becomes faster as $d_{xyz}$ increases, implying that the collective movement decays gradually during the relaxation, and the strength of the original stimulus will affect the regularity of the movement of the atoms.

Since the potential energy $E_p^i$ of the atoms is related to their positions, the $E_p^i$ curves also involve into different groups (Fig. 4), which is in line with the groups described in Fig. 1. The regular fluctuation and the decaying of $E_p^i$ can also be seen in Fig. 4. The regular fluctuation decays faster when the $d_{xyz}$ becomes larger. For weak stimulus, $E_p^i$ fluctuates collectively and regularly, and the curves belonging to different groups could be distinguished readily. While for a stronger stimulus, it is more difficult to distinguish the $E_p^i$ curves belonging to different groups in one shell. For example, when the $d_{xyz}$ = 0.01, group 2a and 2b could be distinguished, while at $d_{xyz}$ = 0.04, they commingle and can no longer be distinguished.

It is interesting to see from Fig. 4 that during the cluster vibrates collectively, when the $E_p^i$ of the atoms in the outer shells (shell-2 and shell-3) becomes lower, $E_p^i$ of the center atom (core-0) becomes higher. A phase difference of $180^o$ between the fluctuations of $E_p^i$ of core-0 and atoms in the outer shells can be found in Fig. 4. However, when $d_{xyz}$ = 0.01 and the $E_p^i$ in the outermost shell (shell-3) decreases to the lowest; the $E_p^i$ in some inner shells (such as shell-1) also increases slowly. The phase difference is very clear when the strength of stimulus is weak ($d_{xyz}$ = 0.01), while it becomes undistinguishable due to the decay of the collective vibration, larger $d_{xyz}$ will make the synchronous fluctuation of $E_p^i$ decay faster.

The calculated results show that the $E_k^i$, the $D^i$ and the $E_p^i$ have different fluctuation characteristics when the excessive energy is released. However, the common point is, the synchronous fluctuation of which decays gradually as the collective vibration of the atoms transforms into irregular thermal movement during the relaxation, and decay could be accelerated by increasing the $d_{xyz}$.

Similar phenomena are also found when the core-3 is disturbed. Result told us that disturbing core-3 will introduce more excessive energy into the cluster and cause the collective vibration of the atoms in the cluster decays faster, indicating that larger disturbing scale will accelerate decaying of the collective vibration of the atoms.

## IV. DISCUSSION

The icosahedral LJ147 cluster is the global optimized configuration of this size, any stimulus of the structure will raise the potential energy of the cluster. Disturbing the inner core will introduce excessive energy into the inner structure, and the following relaxing process will convert the excessive potential energy into kinetic energy of the atoms, which will cause the average kinetic energy of the atoms in the inner shells to be higher than that of atoms in the outer shells. Considering the strength of bonding between atoms in LJ147 cluster distributes spherically from the inner to the outer, the energy of the inner shells diffuses synchronously along the radius of the cluster during the relaxing process, which causes the monopole collective vibrating mode in the cluster. When the whole cluster is disturbed, the excessive energy caused by the stimulus distributes spherically in the cluster, then the released kinetic energy distributes spherically along different radius, which causes the whole cluster expands uniformly along radius, then the collective vibration is built up. From Fig. 2, we could find that, during the relaxation, the amplitude of the fluctuation of $E_k^i$ becomes smaller and finally reaches the state of thermal equilibrium. Then it is reasonable to think that, the conversion from regular monopole vibration to irregular thermal movement is actually the process that the releasing of the excessive energy and the reaching of the thermal equilibrium state.

Fig. 2 displays that stronger stimulus could raises the $E_k^i$ to higher level. According to Salian's theory [15], the decay of monopole collective vibration is caused by the irregular individual movement of atoms, higher energy will increase the possibility of individual movement of atoms, which causes the collective vibration of the cluster decays faster. Stronger and larger-scale original stimulus will introduce more excessive energy into the cluster, so the intensity of the initial kinetic energy will be stronger and the possibility of irregular individual movement of atoms increases, which causes the collective vibration decay faster and provides another proof that high temperature will destroy the collective excitation [15].

From Fig. 3 and Fig. 4, we find that when the distance between the outer shells and the cluster center is shorter, the potential energy of the core-0 is higher. In the previous work [16], we have shown that the outer shells will compress the inner shells, which raises their potential energy; atoms near the cluster center will be compressed more seriously. Since the change of the distance between atoms means the change of potential energy, the constriction of the cluster will raise the potential energy of the atoms in inner shells. When the volume of the cluster constricts to the smallest, the atom of core-0 is compressed to the most degree, which raises the potential energy of the core-0 to the highest level. Because atoms in shell-1 and shell-2 have fewer outer shells, so the compressing effect are not as obvious as core-0, only when the radius of shell-3 constricts to the shortest, could the raising of potential energy of the atoms in shell-1 be noticed.

**V. CONCLUSION**

The monopole collective vibration of LJ147 cluster has been investigated. Some characters of which are noticed as follows:

1. Introducing random dislocation stimulus into the inner cores or the whole cluster, the cluster will show monopole collective vibration in the subsequent relaxation process.

2. The regular collective vibration of the cluster will decay into irregular thermal movement as the cluster reaches to the state of thermal equilibrium.

3. More serious or larger-scale stimulus will cause the monopole collective vibration decay faster to irregular thermal movement.

4. Due to the compress of the outer shells to the inner structure of the cluster, the phase of the fluctuation of the potential energy of the atoms in different shells may be different, the potential energy of atoms in inner shells increases when that of the outer shells decreases.

The results suggest a possible way on how cluster releases the excessive energy induced by a sudden stimulus, which is more like a kind of collective vibration and a subsequent decaying process.

Fig.1 (Color on line) Icosahedral LJ147 cluster denoting its shell structures

Fig. 2 (Color on line) The fluctuation of kinetic energy of atoms after introducing stimulus into core-2. Section A is the initial stage and section B is the final stage.

Fig. 3 (Color on line) The fluctuation of distance between the atoms and the cluster center. Section A is the initial stage and section B is the final stage.

Fig. 4 (Color on line) The fluctuation of potential energy of atoms after introducing stimulus into core-2. Section A is the initial stage and section B is the final stage.

Fig. 1

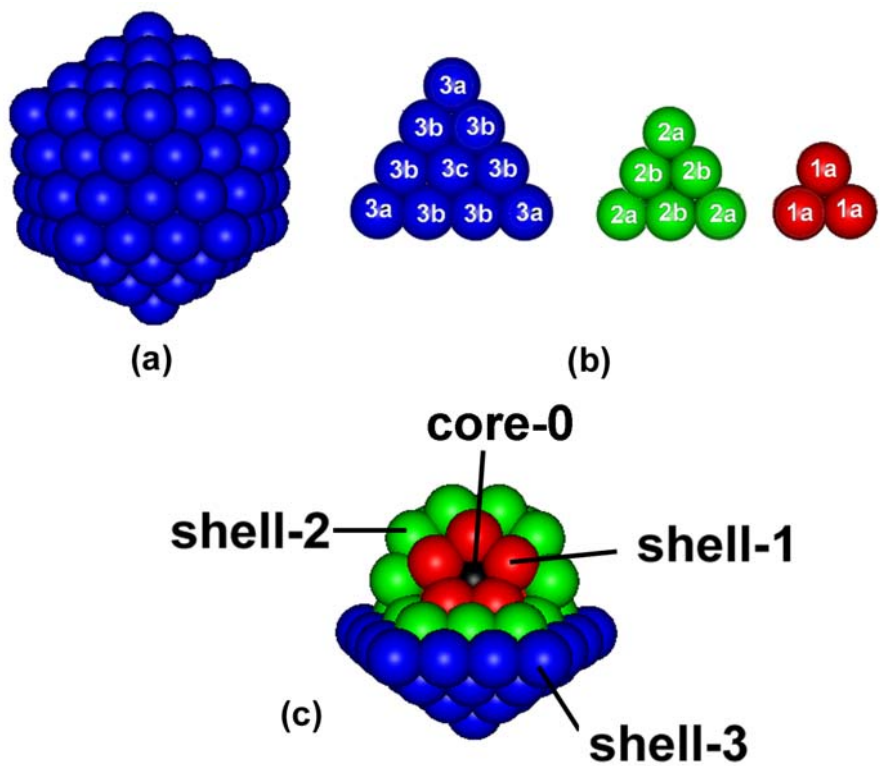

Fig. 2

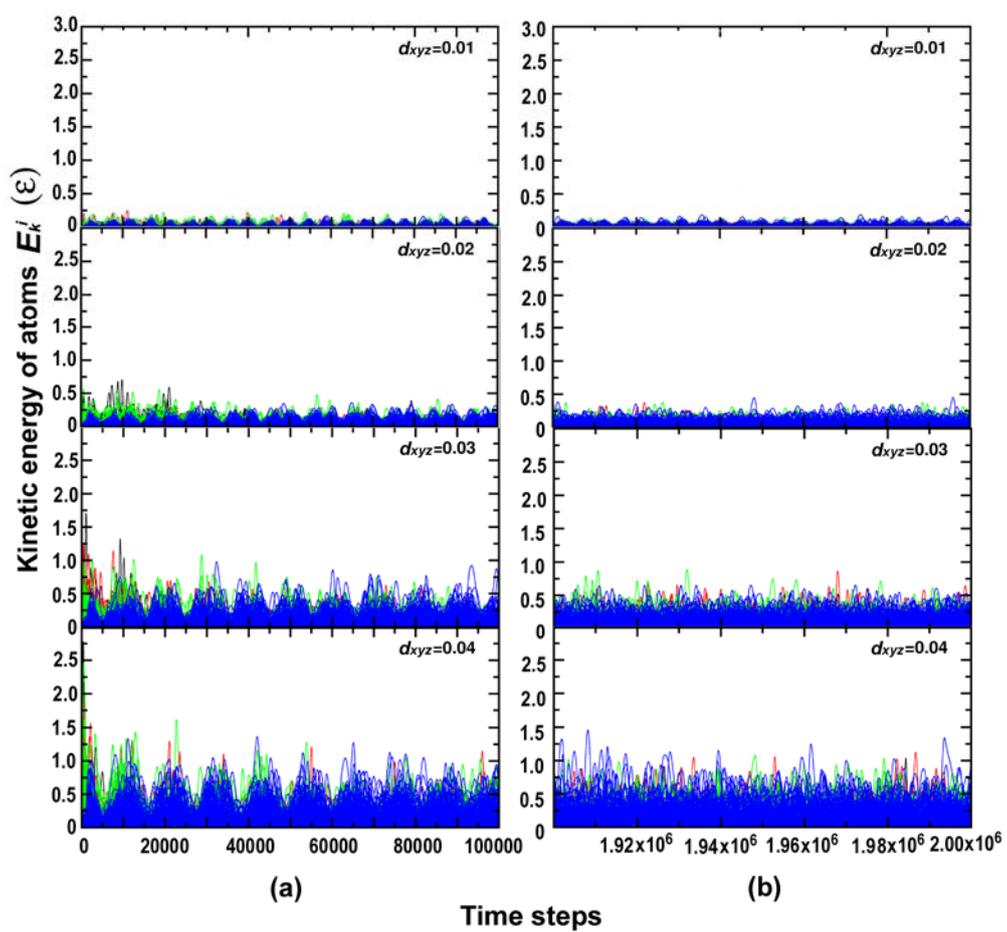

(a)  (b)

Time steps

Fig. 3

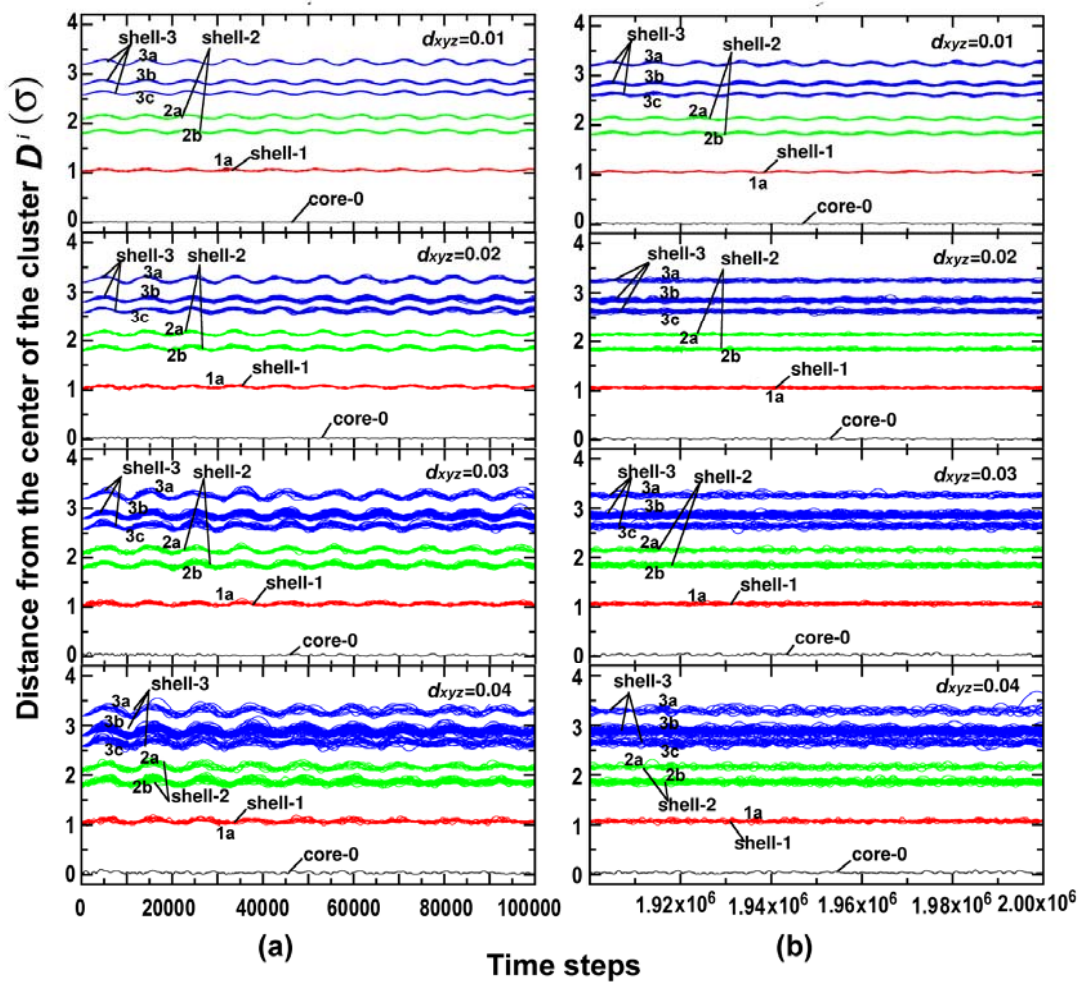

Fig. 4

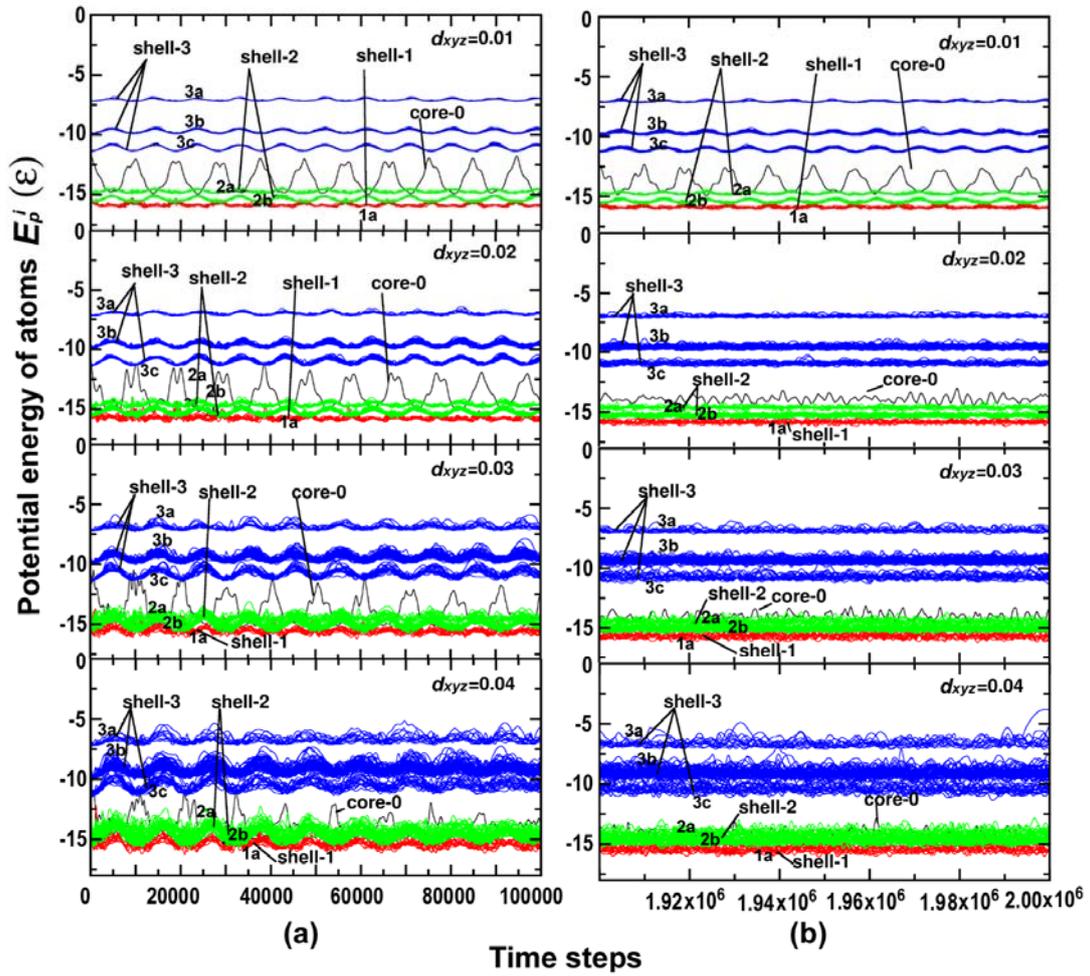